\documentclass[titlepage,12pt]{article}
\usepackage{amsmath}
\usepackage[dvips]{graphicx}
\begin{document}

\title{Bounded solutions of fermions in the background of mixed
vector-scalar P\"{o}schl-Teller-like potentials}
\date{}
\author{L.B. Castro\thanks{%
benito@feg.unesp.br}, A.S. de Castro\thanks{%
castro@pesquisador.cnpq.br} and M.B. Hott\thanks{%
hott@feg.unesp.br} \\
\\
\\
UNESP - Campus de Guaratinguet\'{a}\\
\ \ Departamento de F\'{\i}sica e Qu\'{\i}mica\\
\ \ 12516-410 Guaratinguet\'{a} SP - Brasil\\
\\
}
\date{}
\maketitle

\begin{abstract}
The problem of a fermion subject to a convenient mixing of vector and scalar
potentials in a two-dimensional space-time is mapped into a Sturm-Liouville
problem. For a specific case which gives rise to an exactly solvable
effective modified P\"{o}schl-Teller potential in the Sturm-Liouville
problem, bound-state solutions are found. The behaviour of the upper and
lower components of the Dirac spinor is discussed in detail and some unusual
results are revealed. The Dirac delta potential as a limit of the modified P%
\"{o}schl-Teller potential is also discussed. The problem is also shown to
be mapped into that of massless fermions subject to classical topological
scalar and pseudoscalar potentials.

\bigskip


\end{abstract}

\section{Introduction}

It is well known from the quarkonium phenomenology that the best fit for
meson spectroscopy is found for a mixture of vector and scalar potentials
put by hand in the equations (see, e.g., \cite{luc}). The same can be said
about the treatment of nuclear phenomena describing the influence of the
nuclear medium on the nucleons \cite{ser}. Besides avoiding Klein's paradox,
the spin-orbit splitting is suppressed in hadrons and enhanced in nuclei due
to presence of a scalar potential. The mixed vector-scalar potential has
also been analyzed in 1+1 dimensions for a linear potential \cite{cas2}, as
well as for a general potential which goes to infinity as $|x|\rightarrow
\infty $ \cite{ntd}. In both of those last references it has been concluded
that there is confinement if the scalar coupling is of sufficient intensity
compared with the vector coupling. The problem of a fermion in the
background of an inversely linear potential by considering a proper mixing
of vector and scalar Lorentz structures was re-examined \cite{asc7}. The
problem was mapped into an exactly solvable Sturm-Liouville problem of a Schr%
\"{o}dinger-like equation with an effective Kratzer potential. The case of a
pure scalar potential, already analyzed in \cite{asc6}, was obtained as a
particular case.

Motivated by the success found in references\ \cite{asc7} and \cite{asc6}
the case of massive fermions interacting with background potentials of
vector and scalar natures with a modified P\"{o}schl-Teller-like \cite%
{flugge} configuration is examined here. \ Just a fifty-fifty vector-scalar
mixing is considered in order to make one of the components of the Dirac
spinor to obey a Schr\"{o}dinger equation with an effective modified P\"{o}%
schl-Teller-like potential. The circumstances one can have bound states for
this problem are analyzed and bound-state solutions are obtained. The
behaviour of the eigenenergies as a function of he parameters of the
potential is also taken into account. The impossibility of pair creation by
the vector potential due to the presence of the scalar potential and its
coupling to the fermion mass is remarked. The problem of massive fermions
interacting with delta-like vector and scalar potentials is also discussed
as a limit of the previous problem. Another problem that has also found many
applications in the study of meson physics is that of fermions interacting
with potentials of scalar and pseudoscalar natures, particularly in the
context of the fractionization of the baryon number in the vacuum due to the
topological configurations of the classical backgrounds \cite{jac}. In this
regard it is shown that the problem of massive fermions interacting with
vector and scalar potentials of P\"{o}schl-Teller type can be mapped into
the problem of massless fermions interacting with topological scalar and
pseudoscalar backgrounds, which also admits bound-state solutions. The
fermion bound state for a kink-like pseudoscalar is obtained explicitly and
compared with a similar problem reported previously in the literature \cite%
{wilcz}.

\section{The Dirac equation with mixed vector-scalar potentials in a 1+1
dimension}

In the presence of time-independent vector and scalar potentials the 1+1
dimensional time-independent Dirac equation for a fermion of rest mass $m$
reads

\begin{equation}
H\psi =\left[ c\alpha p+\beta \left( mc^{2}+V_{s}\right) +V_{v}\right] \psi
=E\psi  \label{1a}
\end{equation}

\noindent where $E$ is the energy of the fermion, $c$ is the velocity of
light and $p$ is the momentum operator. $\alpha $ and $\beta $ are Hermitian
square matrices satisfying the relations $\alpha ^{2}=\beta ^{2}=1$ and $%
\left\{ \alpha ,\beta \right\} =0$. The positive definite function $|\psi
|^{2}=\psi ^{\dagger }\psi $, satisfying a continuity equation, is
interpreted as a position probability density and its norm is a constant of
motion. This interpretation is completely satisfactory for single-particle
states \cite{tha}. We use $\alpha =\sigma _{1}$ and $\beta =\sigma _{3}$.
The subscripts for the terms of potential denote their properties under a
Lorentz transformation: $v$ for the time component of the two-vector
potential and $s$ for the scalar potential. Provided that the spinor is
written in terms of the upper and the lower components, $\psi _{+}$ and $%
\psi _{-}$ respectively, \noindent the Dirac equation decomposes into:
\begin{equation}
i\hbar c\psi _{\pm }^{\prime }=\left[ V_{v}-E\mp \left( mc^{2}+V_{s}\right) %
\right] \psi _{\mp }  \label{1c}
\end{equation}

\noindent where the prime denotes differentiation with respect to $x$. In
terms of $\psi _{+}$ and $\psi _{-}$ the spinor is normalized as $%
\int_{-\infty }^{+\infty }dx\left( |\psi _{+}|^{2}+|\psi _{-}|^{2}\right) =1$%
, so that $\psi _{+}$ and $\psi _{-}$ are square integrable functions. It is
clear from the pair of coupled first-order differential eqs. (\ref{1c}) that
both $\psi _{+}$ and $\psi _{-}$ have opposite parities if the Dirac
equation is covariant under $x\rightarrow -x$.

For $V_{v}=0$ and a pure scalar attractive potential, one finds energy
levels for fermions and antifermions arranged symmetrically about $E=0$ \cite%
{cn}. The presence of \ a vector potential breaks this symmetry. For $V_{v}=$
$-V_{s}$ and an attractive scalar potential, the vector potential is
repulsive for fermions but attractive for antifermions. Hence, it can be
anticipated that there is no bound-state solutions for fermions, but only
for antifermions. $V_{v}=$ $+V_{s}$ is the other way around, i.e., the
mixing only holds bound states for fermions. Next this last circumstance is
taken into account.

\section{The P\"{o}schl-Teller-like potential}

Here the particular case $V_{v}=V_{s}$ is considered. The solutions
concerning to $V_{v}=-V_{s}$ can be obtained from the previous one by
applying the charge conjugation operation to the set of eqs. (\ref{1c}),
namely, $\psi _{c}=\sigma _{1}\psi ^{\ast }$, $E\rightarrow -E$, $%
V_{v}\rightarrow -V_{v}$ and $V_{s}\rightarrow V_{s}$. For $V_{v}=V_{s}$ the
lower component is obtained from the knowledge of the upper component
through the equation

\begin{equation}
\psi _{-}=-\,\frac{i\hbar c}{E+mc^{2}}\,~\psi _{+}^{\prime },  \label{psi-}
\end{equation}

\noindent whereas one gets a Sturm-Liouville problem for the upper component

\begin{equation}
-\frac{\hbar ^{2}}{2m}\psi _{+}^{\prime \prime }+\left( \frac{E+mc^{2}}{%
mc^{2}}V_{s}\right) \psi _{+}=\frac{E^{2}-m^{2}c^{4}}{2mc^{2}}~\psi _{+}.
\label{sch}
\end{equation}

\noindent In this way the Dirac problem can be solved by resorting to the
solution of a Schr\"{o}dinger-like problem. Note that eqs. (\ref{psi-}) and (%
\ref{sch}) are valid only for states with $E\neq -mc^{2}$. An inspection of
the first order eqs. (\ref{1c}) reveals that it is impossible to have
normalizable wave functions corresponding to $E=-mc^{2}$. Equation (\ref{sch}%
) is a Schr\"{o}dinger equation with an effective potential $%
V_{eff}=(E+mc^{2})V_{s}/(mc^{2})$ \noindent and an effective energy given by
$E_{eff}=(E^{2}-m^{2}c^{4})/(2mc^{2})$.

For a scalar potential like the symmetric modified P\"{o}schl-Teller \cite%
{flugge} one has

\begin{equation}
V_{s}=-\frac{V_{0}}{\cosh ^{2}\alpha x},\hspace{0.25in}V_{0}>0.  \label{Vs}
\end{equation}

\noindent From eq. (\ref{sch}) it can be noted that the effective energy for
bound states has to be greater than the minimum value of the effective
potential and less than zero, since $V_{s}\rightarrow 0$ for $x\rightarrow
\pm \infty $, that is

\begin{equation}
-(E+mc^{2})V_{0}<\frac{E^{2}-m^{2}c^{4}}{2}<0.
\end{equation}%
The states with $|E|>mc^{2}$ corresponds to the continuous spectrum and the
bound states are to be found in the range $-mc^{2}<E<+mc^{2}$, for $%
V_{0}\geq mc^{2}$. For $V_{0}<mc^{2}$ the bound-state energies are in the
interval $-2V_{0}+mc^{2}<E<+mc^{2}$, and there is an energy gap between $%
-mc^{2}$ and $mc^{2}-2V_{0}$.

The solution of the relativistic problem can be obtained from a comparison
with the nonrelativistic solution of the symmetric modified P\"{o}%
schl-Teller problem treated, for example, in references \cite{flugge}, \cite%
{lan} and \cite{nieto}. The eigenenergies obey the equation

\begin{equation}
\sqrt{m^{2}c^{4}-E^{2}}=\hbar c|\alpha |a_{n},  \label{ener}
\end{equation}

\noindent where $n$ is a natural number, $a_{n}=s-n>0$, and

\begin{equation}
s=\frac{1}{2}\left( -1+\sqrt{1+\frac{8(E+mc^{2})V_{0}}{(\hbar c\alpha )^{2}}}%
\right) .  \label{s}
\end{equation}%
Note that $E>-mc^{2}$ for ensuring that $s>0$.

In Fig.1 it is shown the behaviour of the three lowest states as a function
of $V_{0}$ for $\alpha =1$. For some values of $V_{0}$ it is possible to
have fermion states with negative energy, although they never cross the
value $E=-mc^{2}$, as expected. The bound states are found to the right of
the slanted dashed line (this one is expressed by $-2V_{0}+mc^{2}$) and the
energy gap is the vertical distance between $E=-mc^{2}$ and the slated
dashed line. It can also be seen from this figure that as $V_{0}$ increases
the number of allowed bound states increases.

The behaviour the eigenenergies as a function of $\alpha $ for a fixed value
of $V_{0}$ ($V_{0}<mc^{2}$) has also been analyzed. All the bound-states
energies emerges from the minimum allowed value $E=-2V_{0}+mc^{2}$. As $%
\alpha $ increases the number of allowed bound states decreases, because the
highest level states are ejected from the discrete spectra to the continuum.
That can not be interpreted as a pair production mechanism, since the
fermion bound state becomes a fermion scattering state. The lowest bound
state, $n=0$, is always present and reaches asymptotically the continuum as $%
\alpha \rightarrow \infty $.

\noindent \qquad Following the reference \cite{nieto} one finds that the
upper component of the Dirac spinor can be written in terms of the
Gegenbauer (ultraspherical) polynomial $C_{n}^{\left( a\right) }\left(
z\right) $ (a polynomial of degree $n$),
\begin{equation}
\psi _{+}=N\,\,\left( 1-z^{2}\right) ^{a_{n}/2}C_{n}^{\left(
a_{n}+1/2\right) }\left( z\right) ,\hspace{0.25in}z=\mathrm{tanh}\,\alpha x,
\end{equation}%
where $N$ is a normalization constant. The lower component is obtained from
eq. (\ref{psi-}). By using differential and recurrence relations among the
Gegenbauer polynomials \cite{abr}, one obtains%
\begin{eqnarray}
\psi _{-} &=&N\,\frac{i\hbar c\alpha }{E+mc^{2}}\frac{\left( 1-z^{2}\right)
^{a_{n}/2}}{(2n+2a_{n}+1)}\times  \notag \\
&&\times \left[ (n+a_{n})(n+1)C_{n+1}^{\left( a_{n}+1/2\right)
}(z)-(n+2a_{n})(n+a_{n}+1)C_{n-1}^{\left( a_{n}+1/2\right) }(z)\right] ,
\end{eqnarray}

It is worthwhile to note that by defining $V_{0}\equiv \hbar cg|\alpha |/2$,
where $g>0$, one has that $V_{s}$, defined in eq. (\ref{Vs}), behaves like a
delta potential as $|\alpha |\rightarrow \infty $, that is

\begin{equation}
V_{s}=-\frac{\hbar cg|\alpha |}{2\cosh ^{2}\alpha x}~\underset{|\alpha
|\rightarrow \infty }{\longrightarrow }~-\hbar cg\delta (x).
\end{equation}%
This way the behaviour of a relativistic fermion in the presence of mixed
vector-scalar delta potentials can be studied. In the limit $|\alpha
|\rightarrow \infty $ the parameter $s$ defined in (\ref{s}) can be
approximated for

\begin{equation}
s\underset{|\alpha |\rightarrow \infty }{\longrightarrow }\gamma =\frac{%
E+mc^{2}}{\hbar c|\alpha |}g,
\end{equation}%
and the requirement $\gamma -n>0$ for having bound states is only satisfied
for the lowest state ($n=0$), with eigenenergy satisfying the equation $%
\sqrt{m^{2}c^{4}-E^{2}}=\hbar c|\alpha |\gamma $. Under such conditions the
delta potential has only one bound state with energy $%
E=mc^{2}(1-g^{2})/(1+g^{2})$, which agrees with reference \cite{adame}. The
upper component of the eigenspinor is given by

\begin{equation}
\psi _{+}=N\,\,\left( 1-z^{2}\right) ^{s/2}\underset{|\alpha |\rightarrow
\infty }{\longrightarrow }\psi _{+_{\delta }}=N\,\,e^{-\gamma |\alpha x|},
\end{equation}%
and the lower component is obtained by using (\ref{psi-}), namely

\begin{equation}
\psi _{-_{\delta }}=iN\,g\frac{x}{|x|}\,e^{-\gamma |\alpha x|}.
\end{equation}%
The normalization constant is given by $N=\sqrt{|\alpha |\gamma /(1+g^{2})}$%
. Although the lower component of the eigenspinor is discontinuous at the
origin, the probability density is continuous there. Furthermore, the
nonrelativistic limit can be recovered by taking the strength of the
potential $g$ very small, such that the lower component is suppressed and
the energy is given by $E=mc^{2}(1-2g^{2})$, in consonance with the results
found in reference \cite{diaz}.

\section{The problem of scalar and pseudoscalar potentials}

The problem of fermions in two-dimensions interacting with classical
background potentials has also been considered in quantum field theory to
describe some properties of quasi-one dimensional conductors and some
polymers. One of such an interesting properties is the possibility of
induced fractional fermion number in the vacuum due to interactions of
fermions with topological backgrounds. To illustrate this last phenomenon,
one usually resorts to models of fermions interacting with background
potentials of scalar and/or pseudoscalar natures \cite{jac}. For fermions
interacting with scalar and pseudoscalar potentials, approximate methods
allow\ one to obtain the fractionization of the fermion number, such as in
the model considered in\ \cite{wilcz}, where the fermion is massless and the
scalar, $\phi _{s}$, and pseudoscalar, $\phi _{p}$, potentials obey the
constraint $\phi _{s}^{2}+\phi _{p}^{2}=M^{2}c^{4}$, where $M$ is a constant%
\textrm{.}

A very similar model of massless fermions interacting with scalar and
pseudoscalar potentials can be obtained from this one considered here. In
fact, if one performs a unitary transformation (local chiral rotation) in
eq. (\ref{1a}) $\tilde{H}=UHU^{-1}$ and $\tilde{\psi}=U\psi $, with $U=\exp
\left( -i\sigma _{1}\varphi /2\right) $ and $\varphi ^{\prime
}(x)=-2V_{v}/(\hbar c)$, the transformed Hamiltonian describes a massless
fermion interacting with a scalar $\phi _{s}$ and a pseudoscalar $\phi _{p}$
potential

\begin{equation}
\tilde{H}=\sigma _{1}cp+\sigma _{3}\phi _{s}~+\sigma _{2}\phi _{p},
\end{equation}%
with $\phi _{s}=(mc^{2}-\frac{\hbar c}{2}\varphi ^{\prime })\cos \varphi $
and $\phi _{p}=-(mc^{2}-\frac{\hbar c}{2}\varphi ^{\prime })\sin \varphi $.
One can note that $\phi _{s}^{2}+\phi _{p}^{2}=\left( mc^{2}-\frac{\hbar c}{2%
}\varphi ^{\prime }\right) ^{2}$, which recovers the constraint of the
linear sigma model, with $M^{2}=m^{2}$, as $x\rightarrow \pm \infty $ for $%
\varphi ^{\prime }(x)=2V_{0}/(\hbar c)\cosh ^{2}\alpha x$. One can observe a
kink-like profile for the pseudoscalar field and could use this model to
study the fermion fractionization in the context of quantum field theory. It
is a model which exhibits a trapping of massless neutral fermions by scalar
and pseudoscalar potentials, whose localized states can be obtained by
applying the local unitary transformation to the localized states obtained
in the previous section.

In the section II of the reference \cite{wilcz} the model used to illustrate
the fermion number fractionization via adiabatic approximation, consists of
a pseudoscalar potential, which simulates an infinitely thin soliton and a
uniform scalar potential, which plays the r\^{o}le of the fermion rest mass.
From the case $V_{v}=-\hbar cg\delta (x)$, the bound-state solutions for
those configurations of the scalar and pseudoscalar potentials can be
recovered. In this case one has

\begin{equation}
\varphi (x)=2g~\Theta (x)+k,
\end{equation}%
where $k$ is an constant of integration and $\Theta (x)$ is the Heaviside
step function. With the choice $k=-g+\pi $ the eigenspinor is given by

\begin{equation}
\tilde{\psi}(x)=-N\left(
\begin{array}{c}
\frac{x}{|x|}\left[ \sin (\frac{g}{2})-\frac{1}{g}\cos (\frac{g}{2})\right]
\\
i\left[ \cos (\frac{g}{2})+\frac{1}{g}\sin (\frac{g}{2})\right]%
\end{array}%
\right) \,e^{-\gamma |\alpha x|},
\end{equation}%
which is clearly discontinuous at the origin, although the probability
density is well defined everywhere. Nevertheless, a continuous wave function
at the origin can be obtained by setting $\tan (g/2)=1/g$, namely

\begin{equation}
\tilde{\psi}(x)=-\sqrt{\frac{mc^{2}\sin (g)}{\hbar c}}\left(
\begin{array}{c}
0 \\
i%
\end{array}%
\right) \,\exp \left( -\frac{mc^{2}\sin (g)}{\hbar c}|x|\right) ,
\end{equation}%
with energy given by $E=-mc^{2}\cos (g)$. If one had used the same
representation for the gamma matrices of the reference \cite{wilcz}, the
present solution would coincide with that one found previously, except for
the fact that the pseudoscalar potential in the present case is negative as
well as is the energy for $g>0$, while in Ref. \cite{wilcz} the converse is
true.

\section{Conclusions}

We have succeed in searching for exact Dirac bound solutions for massive
fermions by considering a proper mixing of vector-scalar modified P\"{o}%
schl-Teller-like potentials in 1+1 dimensions. The satisfactory completion
of this task has been possible by means of the methodology of effective
potentials which has transmuted the question into the one of a Schr\"{o}%
dinger-like equation with an effective symmetric modified P\"{o}schl-Teller
potential. We note that the energy levels for particles, emerging from $%
E=mc^{2}$, can have negative values but they never dive into the Dirac sea ($%
E=-mc^{2}$), despite the presence of an electric field. Such a behaviour can
be interpreted as an increase of the effective energy threshold\ to have
pair production, due to the scalar field which couples to the mass of the
fermion.

Recently we have reported on the trapping of neutral fermions by a kink-like
pseudoscalar potential \cite{cashott2}. In this paper it has also been shown
how a model of charged and massive fermions interacting with a convenient
mixing of vector and scalar potentials can be mapped into the one of a
massless neutral fermion interacting with a proper mixing of
scalar-pseudoscalar backgrounds, where the pseudoscalar potential has also a
kink-like profile and the scalar potential a lump-like profile. As a
consequence of the mapping, the bound states of this last problem can also
be obtained from the previous one.

From the point of view of second quantization the problem of fermions under
the action of scalar and pseudoscalar backgrounds illustrates the phenomenon
of the fermion number fractionization which plays a very important r\^{o}le
in low-dimensional systems as well as in particle physics \cite{nie},
although that analysis was not carried out here. For that we need to know
the scattering states in the presence of the scalar and pseudoscalar
potentials. Nevertheless, we recover the bound-state when the pseudoscalar
potential exhibits an infinitely thin solitonic profile. A problem that was
treated in the reference \cite{wilcz} and for which the scattering states
can be easily found.

\bigskip

\noindent {\textbf{Acknowledgments} }

This work was supported in part by means of funds provided by CAPES, CNPq
and FAPESP. We also thank the anonymous referees for valuable comments and
suggestions.

\newpage

\begin{figure}[th]
\begin{center}
\includegraphics[width=9cm, angle=270]{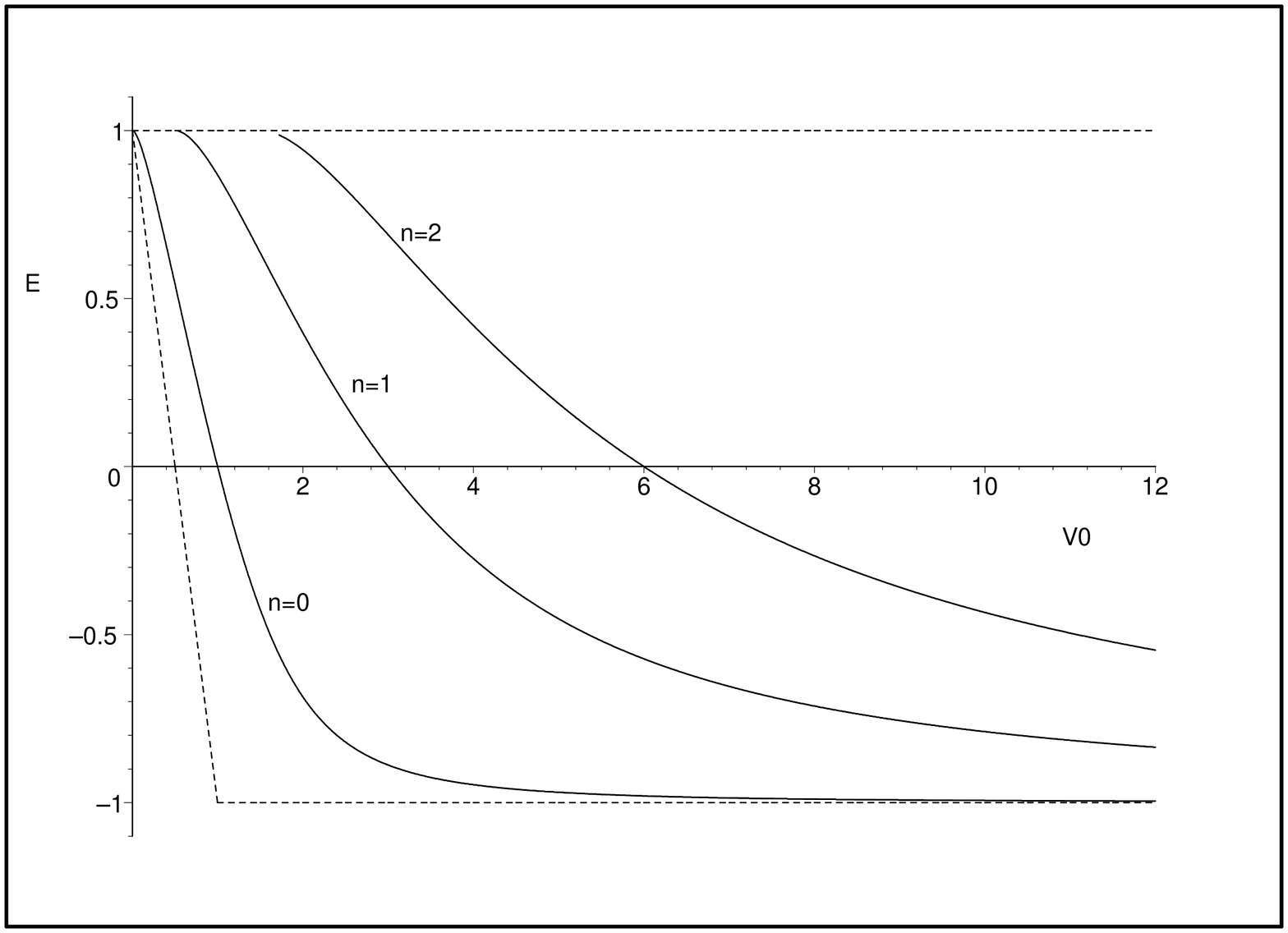}
\end{center}
\par
\vspace*{-0.1cm}
\caption{Dirac eigenvalues for the three lowest energy levels as a function
of $V_{0}$. The horizontal dashed lines stand for $|E|=mc^{2}$ and the
slanted dashed line for $-2V_{0}+mc^{2}$ ($m=\hbar =c=\protect\alpha =1$). }
\label{Fig1}
\end{figure}

\end{document}